\documentclass{optica-article}

\journal{opticajournal} 

\articletype{Research Article}

\usepackage{lineno}

\begin{document}

\title{Derivation and analysis of power offset in fiber-longitudinal power profile estimation using pre-FEC hard-decision data}

\author{Du Tang,\authormark{1} Yingjie Jiang,\authormark{1} Ji Luo,\authormark{2} Yu Chen,\authormark{2} Bofang Zheng,\authormark{2} and Yaojun Qiao\authormark{1,*}}

\address{\authormark{1}The State Key Laboratory of Information Photonics and Optical Communications, School of Information and Communication Engineering, Beijing University of Posts and Telecommunications, Beijing 100876, China\\
\authormark{2}B\&P Laboratory, Huawei Technologies Co. Ltd., Shenzhen 518129, China\\
}

\email{\authormark{*}qiao@bupt.edu.cn} 


\begin{abstract*} 
Utilizing the precise reference waveform regenerated by post-forward error correction (FEC) data, the fiber-longitudinal power profile estimation based on the minimum-mean-square-error method (MMSE-PPE) has been validated as an effective tool for absolute power monitoring. However, when post-FEC data is unavailable, it becomes necessary to rely on pre-FEC hard-decision data, which inevitably introduces hard-decision errors. These hard-decision errors will result in a power offset that undermines the accuracy of absolute power monitoring. In this paper, we present the first analytical expression for power offset in MMSE-PPE when using pre-FEC hard-decision data, achieved by introducing a virtual hard-decision nonlinear perturbation term. Based on this analytical expression, we also establish the first nonlinear relationship between the power offset and the symbol error rate (SER) of $M$-ary quadrature amplitude modulation ($M$-QAM) formats based on Gaussian assumptions. Verified in a numerical 130-GBaud single-wavelength coherent optical fiber transmission system, the correctness of the analytical expression of power offset has been confirmed with 4-QAM, 16-QAM, and 64-QAM formats under different SER situations. Furthermore, the nonlinear relationship between the power offset and SER of $M$-QAM formats has also been thoroughly validated under both linear scale (measured in mW) and logarithmic scale (measured in dB). These theoretical insights offer significant contributions to the design of potential power offset mitigation strategies in MMSE-PPE, thereby enhancing its real-time application. 

\end{abstract*}

\section{Introduction}
Digital signal processing (DSP)-based fiber-longitudinal power profile estimation (PPE) has recently garnered considerable interest and attention from researchers, becoming a prominent topic in optical performance monitoring (OPM) \cite{OPM_DT,OPM2}. Unlike conventional OPM methods, which monitor only cumulative parameters \cite{ONSR1_NL1,ONSR2_NL2,CD1,PMD1,NL1,FOE1} or require additional hardware such as optical time-domain reflectometry (OTDR), PPE reconstructs the signal's power evolution within the fiber link using solely communication data and receiver-side DSP, aided by a first-order nonlinear perturbation model \cite{RP1}. This capability allows for the monitoring of span-wise power evolution \cite{T3,T4}, fiber loss and power anomalies \cite{T2,T6}, fiber type distinctions \cite{fiber_type2,OFC2024_2}, optical amplifier gain spectra \cite{gain,gain2}, passband narrowing \cite{narrowing}, polarization dependent loss (PDL) \cite{PDL1,PDL2}, multi-path interference (MPI) \cite{MPI}, differential group delay (DGD) \cite{DGD_PPE}, inter-band stimulated Raman scattering-induced power transitions \cite{SRS_PPE}, and nonlinear signal-to-noise ratio (SNR) \cite{SNR_PPE}. These diverse applications underscore PPE as a crucial technique in intelligent optical networks with integrated sensing and communication features.

Recently, apart from the applications of PPE, there has been an increasing number of studies looking back on the theoretical performance limitations of PPE. \cite{T7} presented analytical expressions for the two major methods of PPE, the correlation method (CM) and the minimum-mean-square-error (MMSE) method, and obtained a closed-form formula for the spatial resolution of CM. \cite{T7} also confirmed that MMSE approaches the true absolute power profile under a fine spatial step size but is more vulnerable to noise while CM only outputs the relative power profile but has a better noise robustness. Moreover, \cite{T10} quantitatively discussed the ill-posedness problem of MMSE-based PPE (MMSE-PPE) by evaluating the condition number of the nonlinear perturbation matrix, which more fully revealed the performance limitations of MMSE-PPE. Then, to balance the advantages of CM and MMSE, \cite{T5} proposed a generalized model of MMSE-PPE by introducing the Tikhonov regularization, which achieves a trade-off between noise robustness and power sensitivity. Furthermore, to fully understand the PPE behavior under stochastic noise, our recent work \cite{TD} introduced the first analytical noise power expressions for CM-based PPE (CM-PPE) and MMSE-PPE. By defining a new metric called the profile-to-noise ratio (PNR), we have analytically verified the noise tolerance and profile fidelity characteristics of the two methods. All the mentioned theoretical studies of PPE enrich the academic discourse on PPE, paving the way for practical, real-time PPE applications.

However, several limiting factors pose challenges for the practical realization of PPE. One primary limiting factor is the computational complexity and memory limitation and there has been research into developing PPE techniques with lower complexity\cite{JYJ,TD,JYJ2}.Another significant limiting factor is the quality of the received signal. Many works have utilized a nearly perfect regenerated reference waveform using post-forward error correction (FEC) data. However, this may not always be feasible: the post-FEC data might be inaccessible on the DSP chip due to manufacturer restrictions, or even if an open interface for post-FEC data exists, using it to regenerate the reference waveform requires an additional encoding procedure to ensure the regenerated waveform matches the transmitter's, inevitably adding further complexity. If post-FEC data is unavailable, using pre-FEC hard decision (HD) data is an option. However, as revealed in \cite{T8}, using pre-FEC HD data will introduce hard decision errors, which will generate a power offset, impairing the capability for absolute power monitoring. Until now, there has been no clear analytical explanation for the generation of power offset using pre-FEC HD data.

In this paper, to fully understand the reason for the generation of power offset in PPE using pre-FEC HD data, we present the first comprehensive derivation and analysis of the power offset in MMSE-PPE. The main contributions can be summarized as follows:

1) An analytical expression for the power offset of MMSE-PPE using pre-FEC HD data is derived by introducing a virtual HD nonlinear perturbation term. This is the first analytical expression of the power offset, and it is thoroughly verified through numerical simulations.

2) A nonlinear relationship between the power offset and the symbol error rate (SER) of $M$-ary quadrature amplitude modulation ($M$-QAM) formats is established based on Gaussian assumptions. This relationship has also been thoroughly validated under both linear scale (measured in mW) and logarithmic scale (measured in dB) through extensive numerical simulations, providing insights for designing possible power offset mitigation in MMSE-PPE using pre-FEC data.

The paper is organized as follows. Sec. 2 provides the theoretical derivation and analysis of the power offset in MMSE-PPE using pre-FEC HD data. Based on the analytical expression for the power offset, Sec. 3 establishes a nonlinear relationship between the power offset and the SER of $M$-QAM modulation formats based on Gaussian assumptions. Sec. 4 is dedicated to the quantitative validation of the theoretical analyses presented in Sec. 2 and 3, employing numerical simulations for this purpose. Sec. 5 discussed possible power offset mitigation methods based on the analytical expression in Sec. 3. Finally, Sec. 6 presents the conclusions.

\section{The power offset in MMSE-PPE using pre-FEC HD data}

\subsection{Overview of MMSE-PPE}

The primary objective of PPE is to estimate the optical signal's power $P(z)$ at various positions $z$ along the fiber link. $P(z)$ is also referred to as the fiber-longitudinal power evolution. This power evolution can be reflected in the nonlinear Schrödinger equation (NLSE) which characterizes the transmission of optical signals through the fiber:
\begin{equation}
\frac{\partial A(z,t)}{\partial z}+j \frac{\beta_2(z)}{2} \frac{\partial^2 A(z,t)}{\partial t^2}=j \gamma ^{\prime}(z)|A(z,t)|^2 A(z,t)
\end{equation}
where $A(z,t)$ denotes the optical field complex envelope at position $z$ and time $t$ with normalized power, and $\beta_2$ denotes the group velocity dispersion. $P(z)$ is reflected in $\gamma^{\prime}(z)$ which is expressed as:
\begin{equation}
\gamma^{\prime}(z) = \gamma(z) P(0) \exp \left(-\int_0^z \alpha\left(z^{\prime}\right) d z^{\prime}\right)=\gamma(z) P(z)
\end{equation}
where $\gamma(z)$, $P(0)$, and $\alpha(z)$ denote the nonlinear coefficient, the launch power, and the fiber loss, respectively.

To estimate $\gamma^{\prime}(z)$, the first-order nonlinear perturbation model is employed. Consequently, the received signal after transmission through a fiber link of length $L$ can be expressed as follows:
\begin{equation}
A(L, t) \approx U(L, t)+\Delta U(L, t)
\end{equation}
where $U(L,t)$ denotes the received signal affected solely by dispersion (linear distortion) and $\Delta U(L, t)$ represents the nonlinear perturbation term, which is pivotal for PPE. Here, we define $\widehat{D}_{z_1z_2} = \mathcal{F}^{-1}D_{z_1 z_2}(\omega) \mathcal{F}$ as the linear operator, where $\mathcal{F}$ represents the Fourier transform, and $D_{z_1 z_2}(\omega)=\mathrm{exp}(-j\frac{\omega^2}{2}\int_{z_1}^{z_2}\beta_2(z)dz)$. Consequently, $U(L,t)$ can be further expressed as
\begin{equation}
U(L, t) = \widehat{D}_{0 L}[A(0, t)]
\end{equation}

Then, by defining the nonlinear operator $\widehat{N}=|\cdot|^2(\cdot)$, $\Delta U(L, t)$ can be further expressed as:
\begin{equation}
\Delta U(L,t) = \int_0^L \gamma^{\prime}(z) \Delta u_z(L,t) d z 
\end{equation}
where 
\begin{equation}
\Delta u_z(L, t) = j \widehat{D}_{z L}[ \widehat{N}[ \widehat{D}_{0 z} [A(0, t)]]]
\end{equation}

According to \cite{TD}, the discrete spatial version of Eq. (5) with spatial resolution $\Delta z$ (divisible by $L$)  can be written as
\begin{equation}
\Delta U(t)=g(t)\boldsymbol{\gamma^{\prime}}
\end{equation}
where
\begin{equation}
g(t) = [\Delta u_0(t), \Delta u_{\Delta z}(t), ... , \Delta u_{L-\Delta z}(t)]
\end{equation}
\begin{equation}
\boldsymbol{\gamma^{\prime}} = [\gamma^{\prime}(0), \gamma^{\prime}(\Delta z), ... , \gamma^{\prime}(L-\Delta z)]^T
\end{equation}
with $\Delta u_{z}(t) = \Delta u_{z}(L,t)\Delta z$ for simplicity. The superscript $T$ represents the matrix transpose. $g(t)$ is the discrete-spatial normalized perturbation vector at time $t$. For discrete-time signals with a sampling period $T$ and $n+1$ samples, the nonlinear perturbation term can be written as a vector form
\begin{equation}
\boldsymbol{\Delta U} = [\Delta U(0), \Delta U(T), ... , \Delta U(nT)]^T = G\boldsymbol{\gamma^{\prime}}
\end{equation}
where $G$ is a perturbation matrix written as
\begin{equation}
\begin{aligned}
G&=[g(0)^T, g(T)^T, ... ,g(nT)^T]^T\\
&=\left[\begin{array}{cccc}
\Delta u_0(0) & \Delta u_{\Delta z}(0) & \hdots & \Delta u_{L-\Delta z}(0) \\
\Delta u_0(T) & \Delta u_{\Delta z}(T) & \hdots & \Delta u_{L-\Delta z}(T)\\
\vdots & \vdots & \ddots & \vdots\\
\Delta u_0(n T) & \Delta u_{\Delta z}(n T)& \hdots & \Delta u_{L-\Delta z}(nT)
\end{array}\right]
\end{aligned}
\end{equation}

At the receiver, the only two known signals for PPE are the received signal
\begin{equation}
\boldsymbol{A_L}=[A(L,0), A(L,T), ... , A(L,nT)]^T
\end{equation}
and the reference transmitted signal
\begin{equation}
\boldsymbol{A_0}=[A(0,0), A(0,T), ... , A(0,nT)]^T  
\end{equation}
Then, the nonlinear perturbation vector can be obtained by
\begin{equation}
\boldsymbol{\Delta U} = \boldsymbol{A_L} - \widehat{D}_{0 L}[\boldsymbol{A_0}] = \boldsymbol{A_L} - \boldsymbol{U}
\end{equation}
where $\boldsymbol{U} = \widehat{D}_{0 L}[\boldsymbol{A_0}]$. The perturbation matrix $G$ can be calculated according to Eq. (6), (8), and (11) based on the reference transmitted signal $\boldsymbol{A_0}$. 

Finally, after getting $\boldsymbol{\Delta U}$ and $G$, MMSE-PPE estimates $\gamma^{\prime}$ by solving a classical least-squares problem \cite{T3,T7} and the estimated power profile is expressed as
\begin{equation}
\boldsymbol{\widehat{\gamma^{\prime}}}=\left(G^{\dagger} G\right)^{-1} G^{\dagger} \boldsymbol{\Delta U} = \left(G^{\dagger} G\right)^{-1} G^{\dagger} G\boldsymbol{\gamma^{\prime}} = \boldsymbol{\gamma^{\prime}}
\end{equation}

At this juncture, we have reviewed the principle of MMSE-PPE. In the following two subsections, we will discuss the MMSE-PPE using ideal transmitted (post-FEC) data and pre-FEC HD data. 

\subsection{MMSE-PPE using ideal transmitted (post-FEC) data}

If post-FEC data is available, the reference transmitted signal can be regenerated by re-encoding the post-FEC data, as shown in Fig. 1 (a). This reference transmitted signal can be considered identical to the exact transmitted signal at the transmitter. The equations in Sec. 2.1 are already based on the transmitted signal at the transmitter. To distinguish them from those using pre-FEC HD data in the next subsection, we add the subscript 'tx' to $A(0,t)$, resulting in $A_{tx}(0,t) = A(0,t)$. Consequently, some of the equations and variables in Sec. 2.1 can be rewritten as
\begin{equation}
A(L, t) \approx U_{tx}(L, t)+\Delta U_{tx}(L, t)
\end{equation}
\begin{equation}
U_{tx}(L, t) = \widehat{D}_{0 L}[A_{tx}(0, t)]
\end{equation}
\begin{equation}
\Delta U_{tx}(L,t) = \int_0^L \gamma^{\prime}(z) \Delta u_{tx,z}(L,t) d z 
\end{equation}
\begin{equation}
\Delta u_{tx,z}(L, t) = j \widehat{D}_{z L}[ \widehat{N}[ \widehat{D}_{0 z} [A_{tx}(0, t)]]]
\end{equation}
\begin{equation}
\Delta u_{tx, z}(t) = \Delta u_{tx,z}(L,t)\Delta z
\end{equation}
\begin{equation}
g_{tx}(t) = [\Delta u_{tx,0}(t), \Delta u_{tx,\Delta z}(t), ... , \Delta u_{tx,L-\Delta z}(t)]
\end{equation}
\begin{equation}
\Delta U_{tx}(t)=g_{tx}(t)\boldsymbol{\gamma^{\prime}}
\end{equation}
\begin{equation}
G_{tx}=[g_{tx}(0)^T, g_{tx}(T)^T, ... ,g_{tx}(nT)^T]^T
\end{equation}
\begin{equation}
\boldsymbol{A_{tx,0}}=[A_{tx}(0,0), A_{tx}(0,T), ... , A_{tx}(0,nT)]^T  
\end{equation}
\begin{equation}
\boldsymbol{\Delta U_{tx}} = \boldsymbol{A_L} - \widehat{D}_{0 L}[\boldsymbol{A_{tx,0}}] = \boldsymbol{A_L} - \boldsymbol{U_{tx}} = G_{tx}\boldsymbol{\gamma^{\prime}}
\end{equation}

Finally, the estimated power profile using ideal transmitted (post-FEC) data is expressed as
\begin{equation}
\boldsymbol{\widehat{\gamma^{\prime}}_{tx}}=\left(G_{tx}^{\dagger} G_{tx}\right)^{-1} G_{tx}^{\dagger} \boldsymbol{\Delta U_{tx}} = \left(G^{\dagger} G\right)^{-1} G^{\dagger} \boldsymbol{\Delta U} = \boldsymbol{\gamma^{\prime}}
\end{equation}

\begin{figure}[t!]
\centering
\includegraphics[width=28 pc]{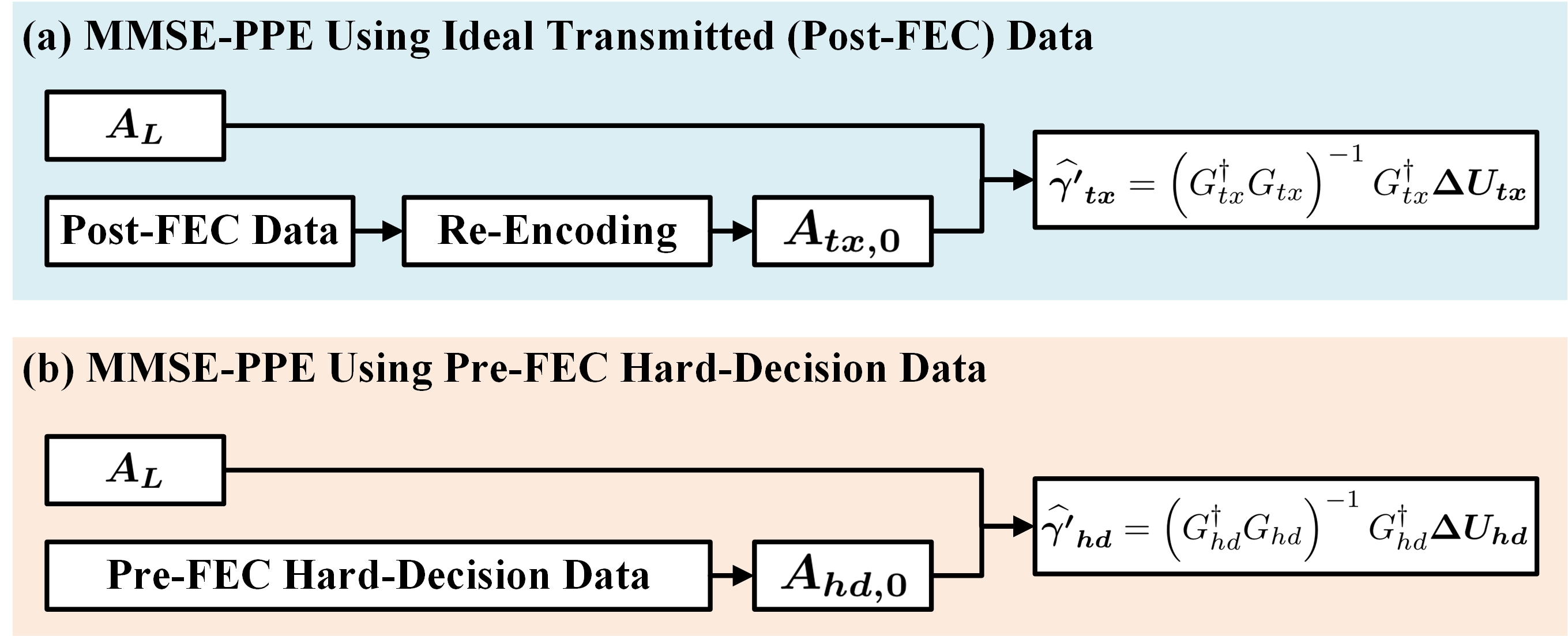}
\caption{The diagram of (a) MMSE-PPE using ideal transmitted (post-FEC) data. (b) MMSE-PPE using pre-FEC HD data.}
\label{fig_2}
\end{figure}

\subsection{MMSE-PPE using pre-FEC HD data}

If post-FEC data is unavailable, using pre-FEC HD data is an alternative, where the reference transmitted signal is regenerated directly from the HD data, as shown in Fig. 1 (b). However, the HD data inevitably differs from the actual transmitted data, resulting in a power offset as mentioned in \cite{T8}. To fully understand the reason for the generation of power offset in PPE using HD data, we re-derive the equations in Sec. 2.1 based on the HD data $A_{hd}(0,t)$.

First, the nonlinear perturbation vector based on the HD data $A_{hd}(0,t)$ can be expressed as
\begin{equation}
\boldsymbol{\Delta U_{hd}} = \boldsymbol{A_L} - \widehat{D}_{0 L}[\boldsymbol{A_{hd,0}}] = \boldsymbol{A_L} - \boldsymbol{U_{hd}}
\end{equation}
where $\boldsymbol{U_{hd}} = \widehat{D}_{0 L}[\boldsymbol{A_{hd,0}}]$.

Then, the perturbation matrix based on the HD data $A_{hd}(0,t)$ can be expressed as
\begin{equation}
G_{hd}=[g_{hd}(0)^T, g_{hd}(T)^T, ... ,g_{hd}(nT)^T]^T
\end{equation}
where
\begin{equation}
g_{hd}(t) = [\Delta u_{hd,0}(t), \Delta u_{hd,\Delta z}(t), ... , \Delta u_{hd,L-\Delta z}(t)]
\end{equation}
\begin{equation}
\Delta u_{hd, z}(t) = \Delta u_{hd, z}(L,t)\Delta z
\end{equation}
\begin{equation}
\Delta u_{hd,z}(L, t) = j \widehat{D}_{z L}[ \widehat{N}[ \widehat{D}_{0 z} [A_{hd}(0, t)]]]
\end{equation}

Finally, the estimated power profile using HD data is expressed as
\begin{equation}
\begin{aligned}
\boldsymbol{\widehat{\gamma^{\prime}}_{hd}}&=\left(G_{hd}^{\dagger} G_{hd}\right)^{-1} G_{hd}^{\dagger} \boldsymbol{\Delta U_{hd}} \\
&= \left(G_{hd}^{\dagger} G_{hd}\right)^{-1} G_{hd}^{\dagger} \left[\boldsymbol{A_L} - \boldsymbol{U_{hd}}\right] \\
&= \left(G_{hd}^{\dagger} G_{hd}\right)^{-1} G_{hd}^{\dagger} \left[\boldsymbol{\Delta U_{tx}} + \boldsymbol{U_{tx,L}} -\boldsymbol{U_{hd,L}}\right]
\end{aligned}
\end{equation}

To obtain the analytical expression of the power offset, we introduce a virtual perturbation vector $\widehat{\boldsymbol{\Delta U_{hd}}}$ which is generated by the HD data transmitted through the same fiber link. Specifically, if we transmit the HD data at the transmitter side, the received virtual signal $\widehat{\boldsymbol{A_{hd,L}}}$ can be written as
\begin{equation}
\widehat{\boldsymbol{A_{hd,L}}} = \boldsymbol{U_{hd}} + \widehat{\boldsymbol{\Delta U_{hd}}}
\end{equation}
where
\begin{equation}
\widehat{\boldsymbol{\Delta U_{hd}}} = [\widehat{\Delta U_{hd}(0)}, \widehat{\Delta U_{hd}(T)}, ... , \widehat{\Delta U_{hd}(nT)}]^T = G_{hd}\boldsymbol{\gamma^{\prime}}
\end{equation}
\begin{equation}
\widehat{\Delta U_{hd}(t)}=g_{hd}(t)\boldsymbol{\gamma^{\prime}}
\end{equation}

Then, for the received virtual signal $\widehat{\boldsymbol{A_{hd,L}}}$, the estimated power profile using MMSE-PPE can be expressed as
\begin{equation}
\boldsymbol{\widehat{\gamma^{\prime}}_{hd,v}}=\left(G_{hd}^{\dagger} G_{hd}\right)^{-1} G_{hd}^{\dagger} \widehat{\boldsymbol{\Delta U_{hd}}} = \boldsymbol{\gamma^{\prime}}
\end{equation}

By substituting Eq. (34) and (36) into (32), we can finally get
\begin{equation}
\begin{aligned}
\boldsymbol{\widehat{\gamma^{\prime}}_{hd}}&=\boldsymbol{\gamma^{\prime}} - \left(G_{hd}^{\dagger} G_{hd}\right)^{-1} G_{hd}^{\dagger} \times \left[\boldsymbol{U_{hd}} - \boldsymbol{U_{tx}} + \widehat{\boldsymbol{\Delta U_{hd}}} - \boldsymbol{\Delta U_{tx}}\right]=\boldsymbol{\gamma^{\prime}} - \mathrm{PO}
\end{aligned}
\end{equation}
where
\begin{equation}
\begin{aligned}
\mathrm{PO} = \left(G_{hd}^{\dagger} G_{hd}\right)^{-1} G_{hd}^{\dagger} \left[\boldsymbol{U_{hd}} - \boldsymbol{U_{tx}} + \widehat{\boldsymbol{\Delta U_{hd}}} - \boldsymbol{\Delta U_{tx}}\right]
\end{aligned}
\end{equation}
is the analytical expression of the power offset. It is evident that the power offset depends not only on the difference between the linearly distorted signals generated by the HD data $\boldsymbol{U_{hd}}$ and the TX data $\boldsymbol{U_{tx}}$, but also depends on the difference between the nonlinear perturbations, $\widehat{\boldsymbol{\Delta U_{hd}}}$ and $\boldsymbol{\Delta U_{tx}}$. The numerical verification of Eq. (38) will be provided in Sec. 4.

\section{The Relationship between power offset and symbol error rate}

To our knowledge, before this work, only \cite{T8} mentioned that the power offset changes with variations in the bit-error rate (BER). However, no exact analytical relationship between the power offset and BER was established in \cite{T8}. This section focuses on determining the relationship between the power offset and the symbol error rate (SER) of $M$-QAM modulation formats. The reason for choosing SER over BER is that SER provides a more intuitive basis for analytical derivations.

\subsection{Assumptions and necessary pre-derivations}

\textbf{Assumption}: the overall noise is modeled as an additive white Gaussian noise (AWGN) with a single-sided power spectral density $N_0$.

Based on this assumption, according to the Eq. 8.3 in \cite{Com}, the SER of $M$-ary amplitude-shift-keying (MASK) with an average symbol energy $E_s$ can be expressed as
\begin{equation}
\mathrm{SER}_{M-\mathrm{ASK},E_s}=2\left(\frac{M-1}{M}\right) Q\left(\sqrt{\frac{6 E_s}{N_0\left(M^2-1\right)}}\right)
\end{equation}
where $Q(\cdot)$ is the Gaussian Q-function expressed as $Q(x)=\int_{x}^{\infty} \frac{1}{\sqrt{2 \pi}} \exp \left(-\frac{y^{2}}{2}\right) d y$. Here, the $Q(\cdot)$ term represents the SER of the symbols at the two edges of the constellation, which can be seen as a conditional probability written as
\begin{equation}
\left.\mathrm{SER}_{M-\mathrm{ASK},E_s}\right|_{edge}=Q\left(\sqrt{\frac{6 E_s}{N_0\left(M^2-1\right)}}\right)
\end{equation}

Hence, the SER of the other symbols can be written as
\begin{equation}
\begin{aligned}
\left.\mathrm{SER}_{M-\mathrm{ASK},E_s}\right|_{\overline{edge}}&=2\left.\mathrm{SER}_{M-\mathrm{ASK},E_s}\right|_{edge}=2Q\left(\sqrt{\frac{6 E_s}{N_0\left(M^2-1\right)}}\right)
\end{aligned}
\end{equation}

Then, for the commonly used $M$-QAM format, SER can be expressed as \cite{Com}
\begin{equation}
\mathrm{SER}_{M-\mathrm{QAM},E_s}=1-\left[1-\mathrm{SER}_{\sqrt{M}-\mathrm{ASK},E_s/2}\right]^2
\end{equation}

Based on Eq. (39) and (41), the relationship between the conditional and the overall SER of $M$-QAM can be obtained:
\begin{equation}
\begin{aligned}
\left.\mathrm{SER}_{\sqrt{M}-\mathrm{ASK},E_s/2}\right|_{edge} = \left(\frac{\sqrt{M}}{2(\sqrt{M}-1)}\right) (1-\sqrt{1-\mathrm{SER}_{M-\mathrm{QAM},E_s}})
\end{aligned}
\end{equation}

Next, to establish the relationship between power offset and SER, we define an error vector $W(t)$ which can be expressed as
\begin{equation}
W(t)=A_{hd}(0, t) - A_{tx}(0, t)
\end{equation}

$W(t)$ is actually a discrete random variable that takes a countable number of possible outcomes, $W(t) \in S$. e.g., for 4-QAM, $S = \{0,2,-2,2j,-2j,2+2j,2-2j,-2+2j,-2-2j\}$.
The probability mass function (PMF) of $W(t)$ can then be expressed as
\begin{equation}
P_W(w)= \begin{cases}\mathrm{Pr}(W=w) & \text { if } w \in S \\ 0 & \text { otherwise }\end{cases}
\end{equation}

Then, according to Bayes' theorem, we get
\begin{equation}
P_{W,A}(w,a) = P_{W|A=a}(w)P_A(a)
\end{equation}
where $A$ can be either $A_{tx}(0, t)$ or $A_{hd}(0, t)$ and $P_A(a)$ denotes the PMF of the modulation format. The conditional PMF $P_{W|A=a}(w)$ of $M$-QAM is related to the conditional SER in Eq. (42). For example, for 4-QAM, 
\begin{equation}
P_{W|A=a}(w)= \begin{cases}\left.\mathrm{SER}_{\sqrt{4}}\right|_{edge}(1-\left.\mathrm{SER}_{\sqrt{4}}\right|_{edge}) & \text { if } w \in S_1 \\ (\left.\mathrm{SER}_{\sqrt{4}}\right|_{edge})^2 & \text { if } w \in S_2 \\ (1-\left.\mathrm{SER}_{\sqrt{4}}\right|_{edge})^2 & \text { if } w = 0 \end{cases}
\end{equation}
where $\left.\mathrm{SER}_{\sqrt{4}}\right|_{edge} = \left.\mathrm{SER}_{\sqrt{4}-\mathrm{ASK},E_s/2}\right|_{edge}$ for simplicity, and $S_1 = \{2,-2,2j,-2j\}$, $S_2 = \{2+2j,2-2j,-2+2j,-2-2j\}$. It is clear that if $w \neq 0$, $P_{W|A=a}(w)$ has a relationship with the square of $\left.\mathrm{SER}_{\sqrt{4}}\right|_{edge}$, which means the relationship between $P_{W|A=a}(w)$ and $\mathrm{SER}_{4-\mathrm{QAM},E_s}$ can be written as:
\begin{equation}
P_{W|A=a}(w)= \begin{cases} \mathrm{SER}_{4-\mathrm{QAM},E_s}+\sqrt{1-\mathrm{SER}_{4-\mathrm{QAM},E_s}}-1 & \text { if } w \in S_1 \\ -\mathrm{SER}_{4-\mathrm{QAM},E_s}-2\sqrt{1-\mathrm{SER}_{4-\mathrm{QAM},E_s}}+2 & \text { if } w \in S_2 
\end{cases}
\end{equation}

For other $M$-QAM constellations, similar relationship applies and without loss of generality, we can get the relationship between $P_{W|A=a}(w)$ and $\mathrm{SER}_{M-\mathrm{QAM},E_s}$:
\begin{equation}
\begin{aligned}
P_{W|A=a}(w) &= k\mathrm{SER}_{M-\mathrm{QAM},E_s} + p\sqrt{1-\mathrm{SER}_{M-\mathrm{QAM},E_s}}+q & \text { if } w \neq 0
\end{aligned}
\end{equation}
where $k$, $p$, and $q$ are modulation-format-related parameters and can be calculated following the above derivations with certain modulation formats.

\subsection{Derivation of the relationship between power offset and symbol error rate}

The power offset expression in Eq. (38) can be divided into two parts: the first part is $\left(G_{hd}^{\dagger} G_{hd}\right)^{-1}$, and the second part is $G_{hd}^{\dagger} \left[\boldsymbol{U_{hd}} - \boldsymbol{U_{tx}} + \widehat{\boldsymbol{\Delta U_{hd}}} - \boldsymbol{\Delta U_{tx}}\right]$.

The first part is the matrix form of the spatial resolution function (SRF) as mentioned in \cite{T7,T8, TD}. If we assume that the nonlinear perturbation terms at different positions are ideally independent, $\left(G_{hd}^{\dagger} G_{hd}\right)^{-1}$ becomes a diagonal matrix and will not affect the second part. Even if this ideal assumption is not satisfied, the first part only acts as a linear convolution function, which will not change the relationship between the variables in the second part.

The second part becomes the key to establishing the relationship between the power offset and SER. This second part can be further divided into two sub-parts: $G_{hd}^{\dagger} \left[\boldsymbol{U_{hd}} - \boldsymbol{U_{tx}} \right]$ and $G_{hd}^{\dagger} \left[\widehat{\boldsymbol{\Delta U_{hd}}} - \boldsymbol{\Delta U_{tx}}\right]$. 

For the first sub-part, each element of $G_{hd}$ is a nonlinear perturbation term $\Delta u_{hd,z}(t)$ and each element of $\boldsymbol{U_{hd}} - \boldsymbol{U_{tx}}$ can be further expressed as
\begin{equation}
\begin{aligned}
U_{hd}(L, t)-U_{tx}(L, t)&=\widehat{D}_{0 L}[A_{hd}(0, t)] - \widehat{D}_{0 L}[A_{tx}(0, t)]\\
&=\widehat{D}_{0 L}[A_{hd}(0, t)-A_{tx}(0, t)]\\
&=\widehat{D}_{0 L}[W(t)]
\end{aligned}
\end{equation}

Then, the first sub-part can be interpreted as an equivalent correlation calculation between $\Delta u_{hd,z}(t)$ and $\widehat{D}_{0 L}[W(t)]$, which can be expressed as
\begin{equation}
\begin{aligned}
E\left[(\Delta u_{hd,z}(t))^* \widehat{D}_{0 L}[W(t)]\right] &=  -j\Delta z E\left[(\widehat{D}_{z L} [\widehat{N}[ \widehat{D}_{0 z} [A_{hd}(0, t)]]])^*\widehat{D}_{z L} [\widehat{D}_{0 z} [W(t)]] \right]\\
& = -j\Delta z \sum_{a} \sum_{w} (\widehat{N}[ \widehat{D}_{0 z} [A_{hd}(0, t)]])^*\widehat{D}_{0 z} [W(t)] \\& \quad \times P_{W|A_{hd}=a}(w)P_{A_{hd}}(a)
\end{aligned}
\end{equation}

It is evident that the first sub-part is a function of $P_{W|A_{hd}=a}(w)P_{A_{hd}}(a)$. For a given $M$-QAM modulation format, $P_{A_{hd}}(a)$ is known and remains independent of the SER. $P_{W|A_{hd}=a}(w)$ has already been expressed in Eq. (49) as a function of SER. Consequently, the relationship between the first sub-part and SER depends on $P_{W|A_{hd}=a}(w)$, which is expressed as $k\mathrm{SER}_{M-\mathrm{QAM},E_s} + p\sqrt{1-\mathrm{SER}_{M-\mathrm{QAM},E_s}}+q$.

The second sub-part, akin to the first sub-part, can also be interpreted as an equivalent correlation computation between $\Delta u_{hd,z}(t)$ and $\widehat{\Delta U_{hd}(t)} - \Delta U_{tx}(t)$. Based on Eq. (30), (31), and (44), $\Delta u_{hd,z}(t)$ can be further derived as follows:
\begin{equation}
\begin{aligned}
\Delta u_{hd,z}(t) & = j \Delta z \widehat{D}_{z L}[ \widehat{N}[ \widehat{D}_{0 z} [A_{hd}(0, t)]]]\\
& = j \Delta z \widehat{D}_{z L}[ \widehat{N}[ \widehat{D}_{0 z} [A_{tx}(0, t)+W(t)]]]\\
& = j \Delta z \widehat{D}_{z L}[ \widehat{N}[ \widehat{D}_{0 z} [A_{tx}(0, t)]]+\widehat{N}[ \widehat{D}_{0 z} [W(t)]] \\
& + 2|\widehat{D}_{0 z}[A_{tx}(0, t)]|^2\widehat{D}_{0 z}[W(t)] + 2|\widehat{D}_{0 z}[W(t)]|^2\widehat{D}_{0 z}[A_{tx}(0, t)] \\
& + (\widehat{D}_{0 z}[A_{tx}(0, t)])^2(\widehat{D}_{0 z}[W(t)])^* + (\widehat{D}_{0 z}[W(t)])^2(\widehat{D}_{0 z}[A_{tx}(0, t)])^*] \\
& = \Delta u_{tx,z}(t) + \overline{\Delta u_{hd,z}(t)}
\end{aligned}
\end{equation}
where $\Delta u_{tx,z}(t) = j \Delta z \widehat{D}_{z L}[ \widehat{N}[ \widehat{D}_{0 z} [A_{tx}(0, t)]]]$ has already been defined in Eq. (20), and $\overline{\Delta u_{hd,z}(t)}$ denotes the remaining terms. 
Subsequently, based on Eq. (21), (22), (29), and (35), $\widehat{\Delta U_{hd}(t)} - \Delta U_{tx}(t)$ can be further derived as
\begin{equation}
\begin{aligned}
\widehat{\Delta U_{hd}(t)} - \Delta U_{tx}(t) &= \sum_{n=0}^{L/\Delta z-1} (\Delta u_{hd,n\Delta z}(t) - \Delta u_{tx,n\Delta z}(t))\gamma^{\prime}(n\Delta z)= \sum_{n=0}^{L/\Delta z-1} \overline{\Delta u_{hd,n\Delta z}(t)}\gamma^{\prime}(n\Delta z)
\end{aligned}
\end{equation}

Based on Eq. (52) and (53), the correlation computation between $\Delta u_{hd,z}(t)$ and $\widehat{\Delta U_{hd}(t)} - \Delta U_{tx}(t)$ can be expressed as
\begin{equation}
\begin{aligned}
&E\left[(\Delta u_{hd,z}(t))^* (\widehat{\Delta U_{hd}(t)} - \Delta U_{tx}(t))\right]\\
&= E\left[(\Delta u_{hd,z}(t))^* \sum_{n=0}^{L/\Delta z-1} \overline{\Delta u_{hd,n\Delta z}(t)}\gamma^{\prime}(n\Delta z) \right] \\
&=  \sum_{n=0}^{L/\Delta z-1} \gamma^{\prime}(n\Delta z) E\left[(\Delta u_{hd,z}(t))^*\overline{\Delta u_{hd,n\Delta z}(t)}\right] \\
&=  \sum_{n=0}^{L/\Delta z-1} \gamma^{\prime}(n\Delta z) \sum_{a}\sum_{w} (\Delta u_{hd,z}(t))^*\overline{\Delta u_{hd,n\Delta z}(t)}\times P_{W|A_{tx}=a}(w)P_{A_{tx}}(a) 
\end{aligned}
\end{equation}
where $\Delta u_{hd,z}(t)$ and $\overline{\Delta u_{hd,z}(t)}$ have already been expressed in Eq. (52), both depending on $A_{tx}(0,t)$ and $W(t)$, thus the $P_{W|A_{tx}=a}(w)P_{A_{tx}}(a)$ can be used here. 

Similar to the first sub-part, the second sub-part is a function of $P_{W|A_{tx}=a}(w)P_{A_{tx}}(a)$. Hence, the relationship between the second sub-part and SER depends on $P_{W|A_{tx}=a}(w)$, which is also expressed as $k\mathrm{SER}_{M-\mathrm{QAM},E_s} + p\sqrt{1-\mathrm{SER}_{M-\mathrm{QAM},E_s}}+q$.

According to the above derivation and analysis, we finally obtain the nonlinear relationship between power offset and SER:
\begin{equation}
\mathrm{po}(z) =  k\mathrm{SER}_{M-\mathrm{QAM},E_s} + p\sqrt{1-\mathrm{SER}_{M-\mathrm{QAM},E_s}}+q 
\end{equation}
where $\mathrm{po}(z)$ denotes the element of $\mathrm{PO}$ at position $z$. The position-dependence of the power offset arises from the term $\gamma^{\prime}(n\Delta z)$ in Eq. (54). Consequently, we can infer that the power offset as a function of position should resemble $\gamma^{\prime}(n\Delta z)$. The numerical verification of Eq. (55) will be provided in Sec. 4.

\section{Numerical verification}

\begin{figure}[b!]
\centering
\includegraphics[width=24 pc]{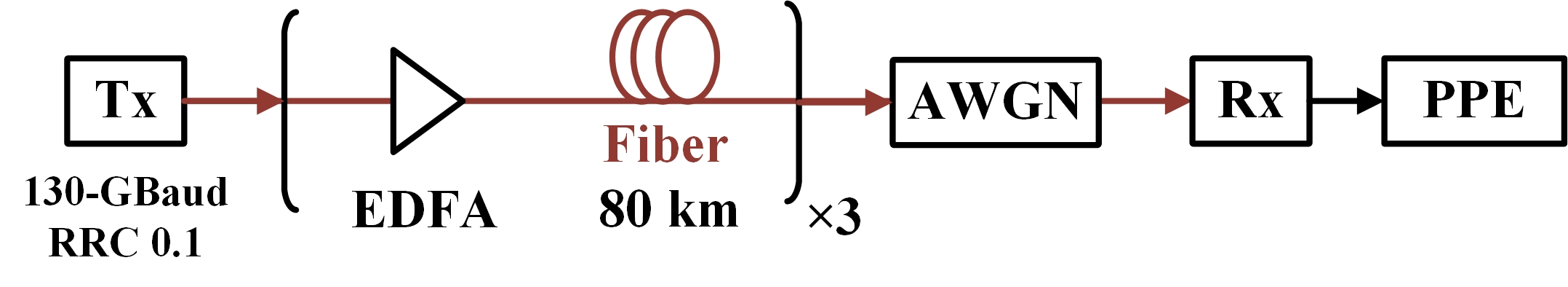}
\caption{Block diagram of the numerical system.}
\label{fig_3}
\end{figure}

In Sec. 2 and 3, we have derived the analytical expression for the power offset of MMSE-PPE using pre-FEC HD data expressed by Eq. (38), and established the relationship between power offset and SER expressed by Eq. (55). In this section, we aim to validate Eq. (38) and Eq. (55) through numerical simulations. As illustrated in Fig. 2, a 130-GBaud single-wavelength coherent optical fiber transmission system is employed. The modulation format is $M$-QAM where $M \in \{ 4, 16, 64 \}$.  A root-raised cosine (RRC) filter with a roll-off factor of 0.1 was applied for pulse shaping. We considered a 3 $\times$ 80 km fiber link and the fiber parameters were consistent with \cite{TD}, with $\alpha = 0.2$ dB/km, $D = 17$ ps/nm/km, and $\gamma = 1.30$ $\text{W}^{-1}\text{km}^{-1}$. The launch power for each span was maintained the same, varying from 5 dBm to 8 dBm in 1 dB intervals. To investigate the power offset under different SERs, we introduced additional AWGN with varying $N_0$ to achieve different SERs, and no other noise was considered in each inline Erbium-doped fiber amplifier (EDFA). To verify the correctness of the derivations, only essential impairments and distortions such as fiber loss, chromatic dispersion, and fiber nonlinearity were considered, aside from the additional AWGN. The DSP at the receiver side was the same as \cite{TD}, including chromatic dispersion compensation, matched filtering, and carrier phase recovery (CPR). The ideal transmitted  (post-FEC) data mentioned in Sec. 2.2 in this numerical system was the same data sent at the transmitter side, while the pre-FEC HD data was the HD data after CPR. Finally, MMSE-PPE was performed based on $\boldsymbol{A_L}$ and $\boldsymbol{A_{hd,0}}$ (or $\boldsymbol{A_{tx,0}}$). 

\begin{figure}[t!]
\centering
\includegraphics[width=21 pc]{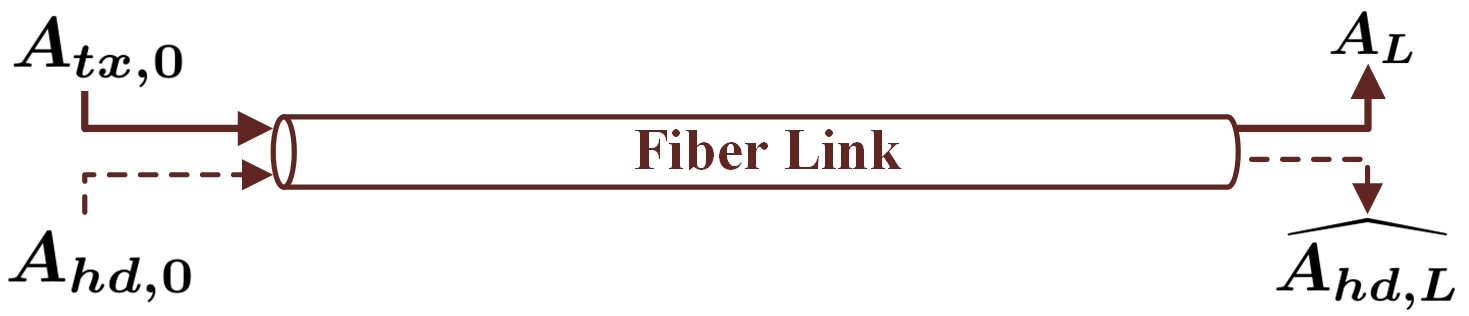}
\caption{Diagram of re-transmitting the HD
data at the transmitter side and obtaining the received signal $\widehat{\boldsymbol{A_{hd,L}}}$.}
\label{fig_4}
\end{figure}

\begin{figure}[b!]
\centering
\includegraphics[width=32 pc]{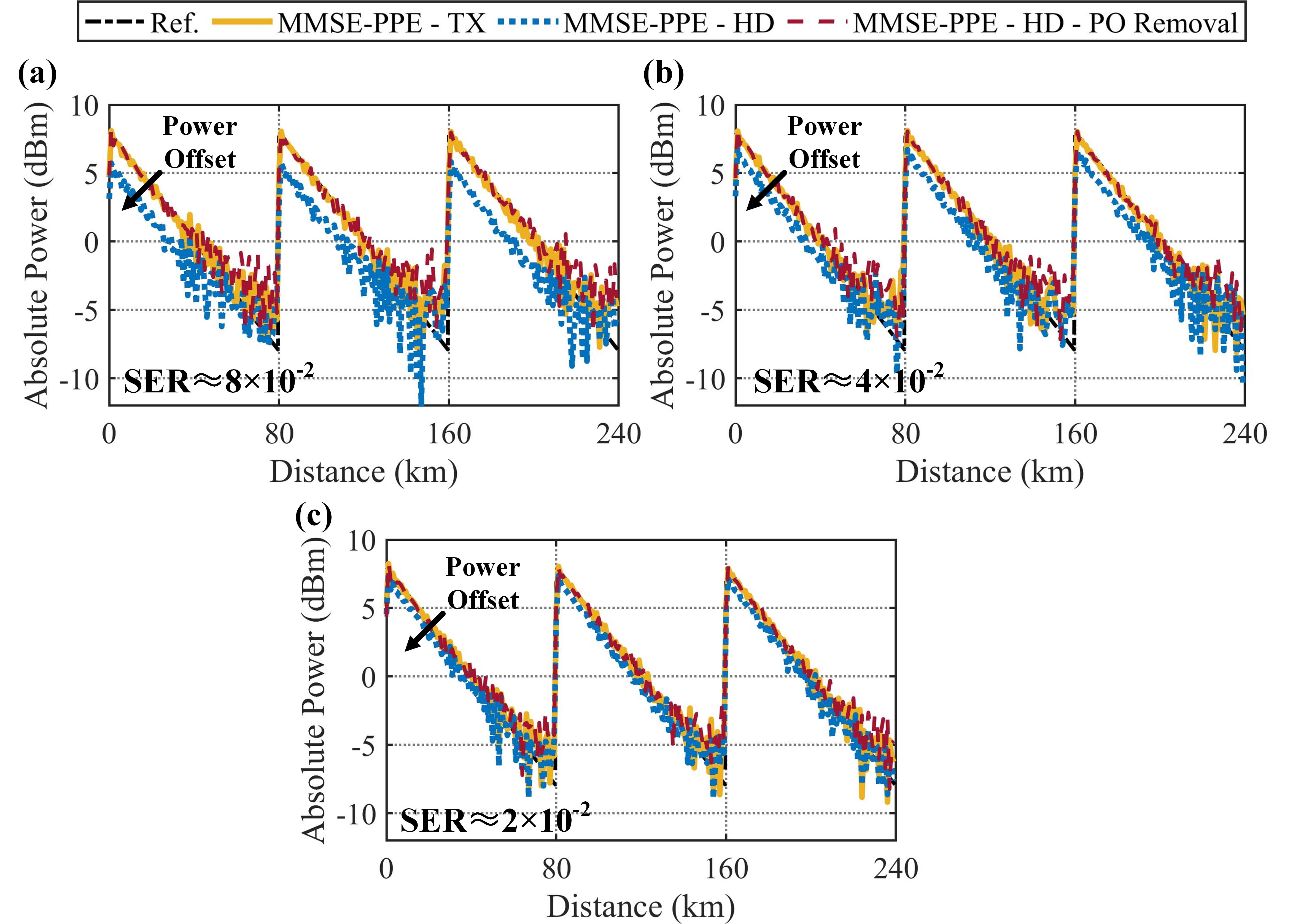}
\caption{The estimated power profiles with MMSE-PPE using ideal transmitted (post-FEC) data, pre-FEC HD data, and pre-FEC HD data with ideal PO removal under different SER cases ((a) $8\times 10^{-2}$, (b) $4 \times 10^{-2}$, and (c) $2 \times 10^{-2}$) of 16-QAM when the launch power is 8 dBm.}
\label{fig_5}
\end{figure}

To verify Eq. (38), the key step is to obtain the virtual nonlinear perturbation $\widehat{\boldsymbol{\Delta U_{hd}}}$. Since the definition of $\widehat{\boldsymbol{\Delta U_{hd}}}$ is the nonlinear perturbation generated by the HD data transmitted through the same fiber link, we re-transmitted the HD data at the transmitter side and obtained the received signal $\widehat{\boldsymbol{A_{hd,L}}} = \boldsymbol{U_{hd}} + \widehat{\boldsymbol{\Delta U_{hd}}}$. Based on Eq. (38), we can then calculate the power offset and remove it from the original power profile, a procedure we term 'ideal PO removal'. To aid readers' understanding, the process described above is depicted in Fig. 3. 

First, we compare the estimated power profiles with different MMSE-PPE schemes with 1-km spatial resolution when the launch power is 8 dBm. As shown in Fig. 4 (a)-(c), compared with the power profiles using ideal transmitted data (yellow dotted lines), power offset is generated by using the HD data (blue lines) for 16-QAM modulation with $8\times 10^{-2}$, $4 \times 10^{-2}$, and $2 \times 10^{-2}$ SER values, respectively. Then, by applying ideal PO removal (red dash-dotted lines), the power profiles using HD data align closely with those using transmitted data, thereby initially verifying the correctness of Eq. (38). Note that no stochastic noise is included in the analytical models and derivations for simplicity. In practice, stochastic noise is inevitable and results in glitches in the estimated power profiles, a phenomenon we have already analyzed in \cite{TD}. Hence, in actual situations, the power offset will be influenced by noise, and the end of each fiber span is more vulnerable to stochastic noise due to the lower nonlinearity level, a phenomenon also observed in many works \cite{T5, T10, TD}. Since this work focuses on the nature of power offset, the impact of noise should be minimized. Therefore, unless otherwise specified, we focus on the beginning of each fiber span due to its higher noise robustness. 

\begin{figure}[b!]
\centering
\includegraphics[width=30 pc]{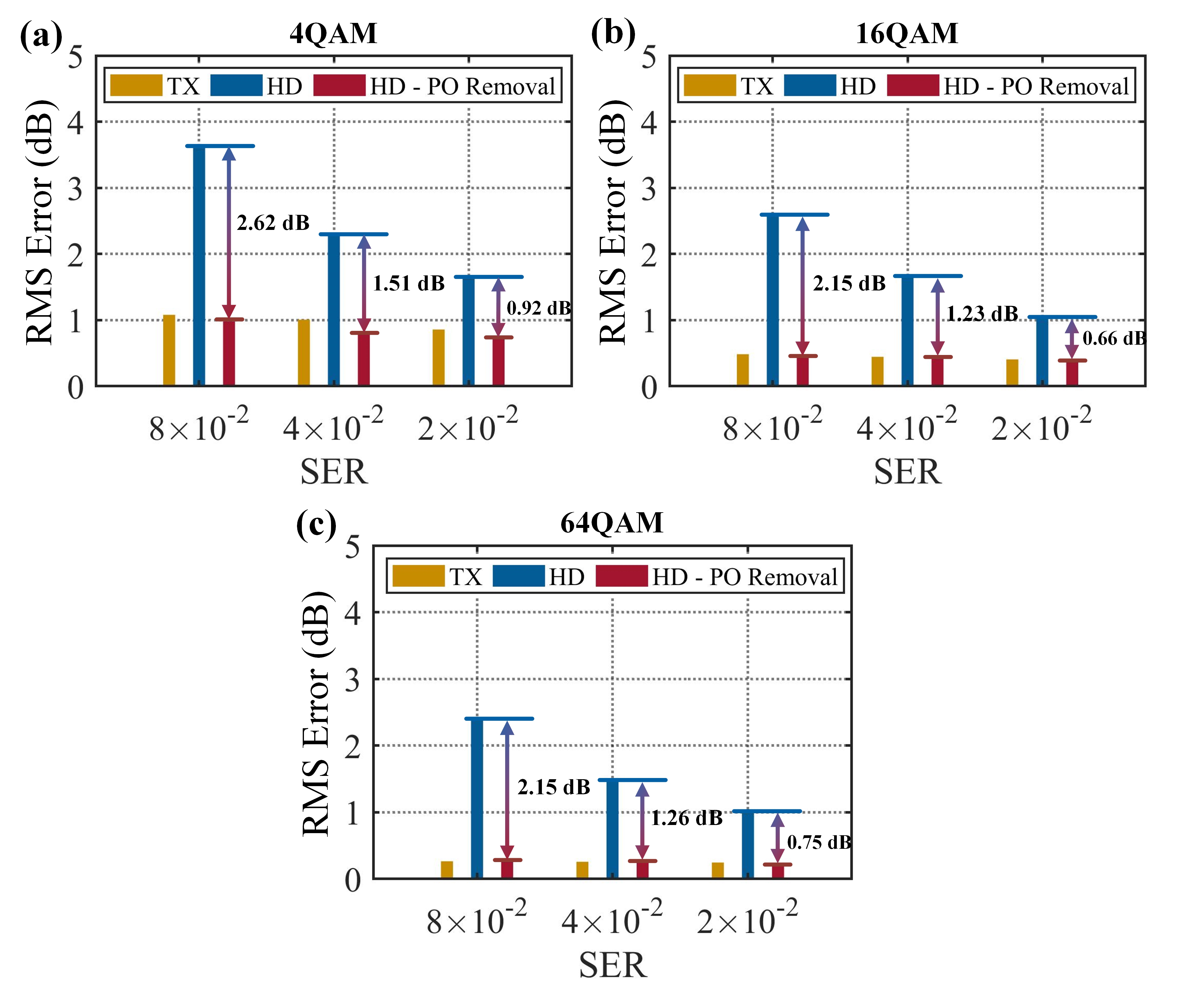}
\caption{The RMS errors between the reference actual power profiles and the power profiles estimated by three types of MMSE-PPE under different SER conditions when the launch power is 8 dBm. (a) 4-QAM. (b) 16-QAM. (c) 64-QAM. }
\label{fig_6}
\end{figure}

Then, we investigated the root mean square (RMS) errors between the reference actual power profiles and the power profiles estimated by three types of MMSE-PPE under different SER conditions when the launch power is 8 dBm: those using ideal transmitted data, those using HD data, and those using HD data but applying ideal PO removal, as shown in Fig. 5. To conduct a comprehensive investigation, three types of $M$-QAM modulation formats with $M \in \{ 4, 16, 64 \}$ are included. The RMS error decreases with the SER for all three MMSE-PPE schemes. For a given SER, the RMS error decreases as $M$ increases. This is because a higher noise level results in a higher SER, and different modulation formats exhibit varying levels of noise robustness. To achieve a given SER, the required noise level for each modulation format differs. Specifically, to achieve an $8\times 10^{-2}$ SER, a higher noise level is applied to 4-QAM compared to 64-QAM, resulting in a higher RMS error for 4-QAM. For a given SER and modulation format, MMSE-PPE using HD data shows the highest RMS error. After using ideal PO removal, the RMS error is significantly reduced, reaching the same level as MMSE-PPE based on ideal transmitted data, further corroborating the correctness of Eq. (38). However, it is important to note that this ideal PO removal is entirely impractical in real-world scenarios since the virtual nonlinear perturbation is unable to be obtained. This method effectively explores the nature of power offset but is not intended to eliminate power offset in actual situations.

\begin{figure}[b!]
\centering
\includegraphics[width=22 pc]{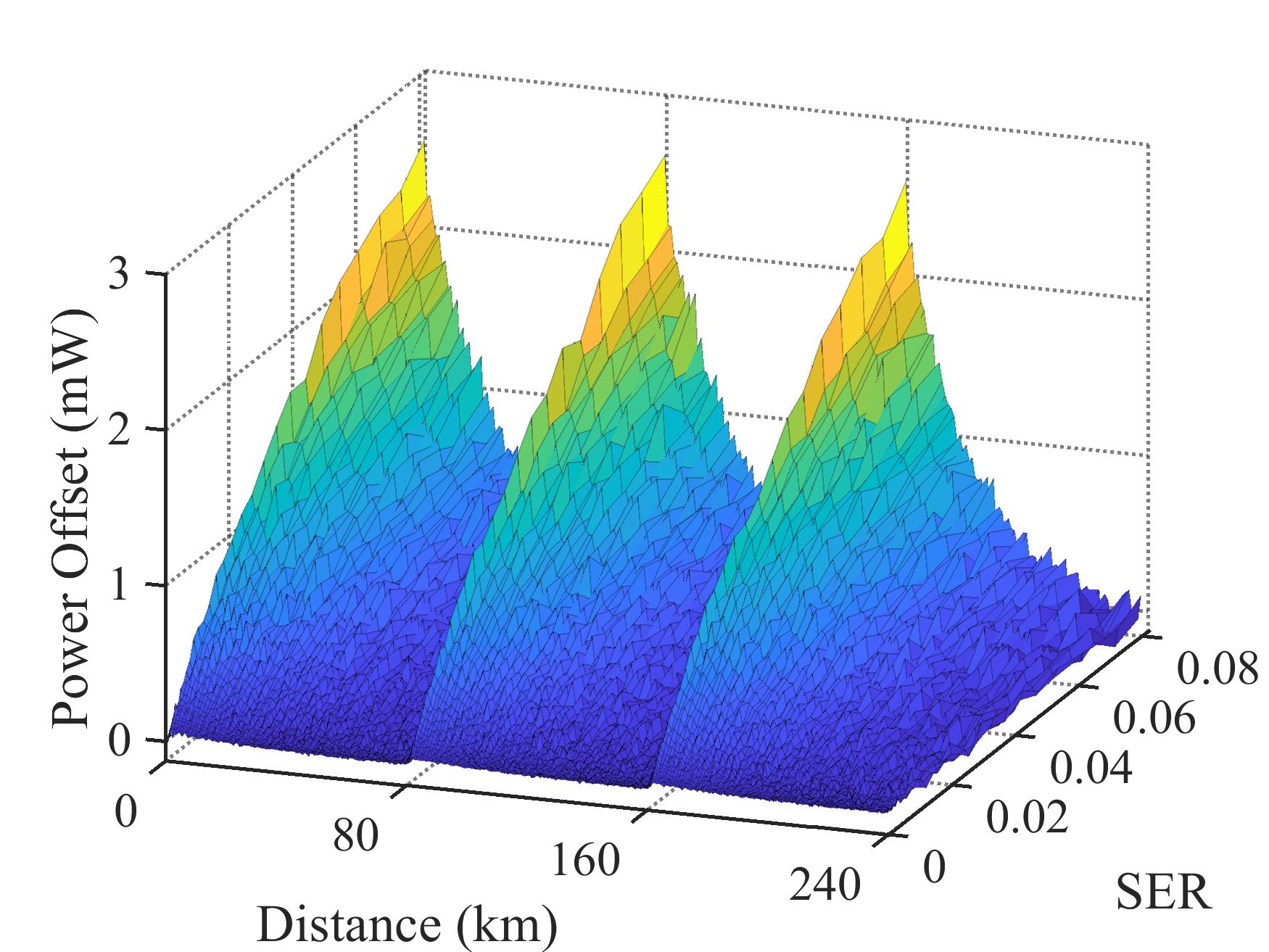}
\caption{The power offset in a unit of mW as a function of transmission distance and SER (16-QAM with 8 dBm launch power).}
\label{fig_7}
\end{figure}

\begin{figure}[t!]
\centering
\includegraphics[width=31 pc]{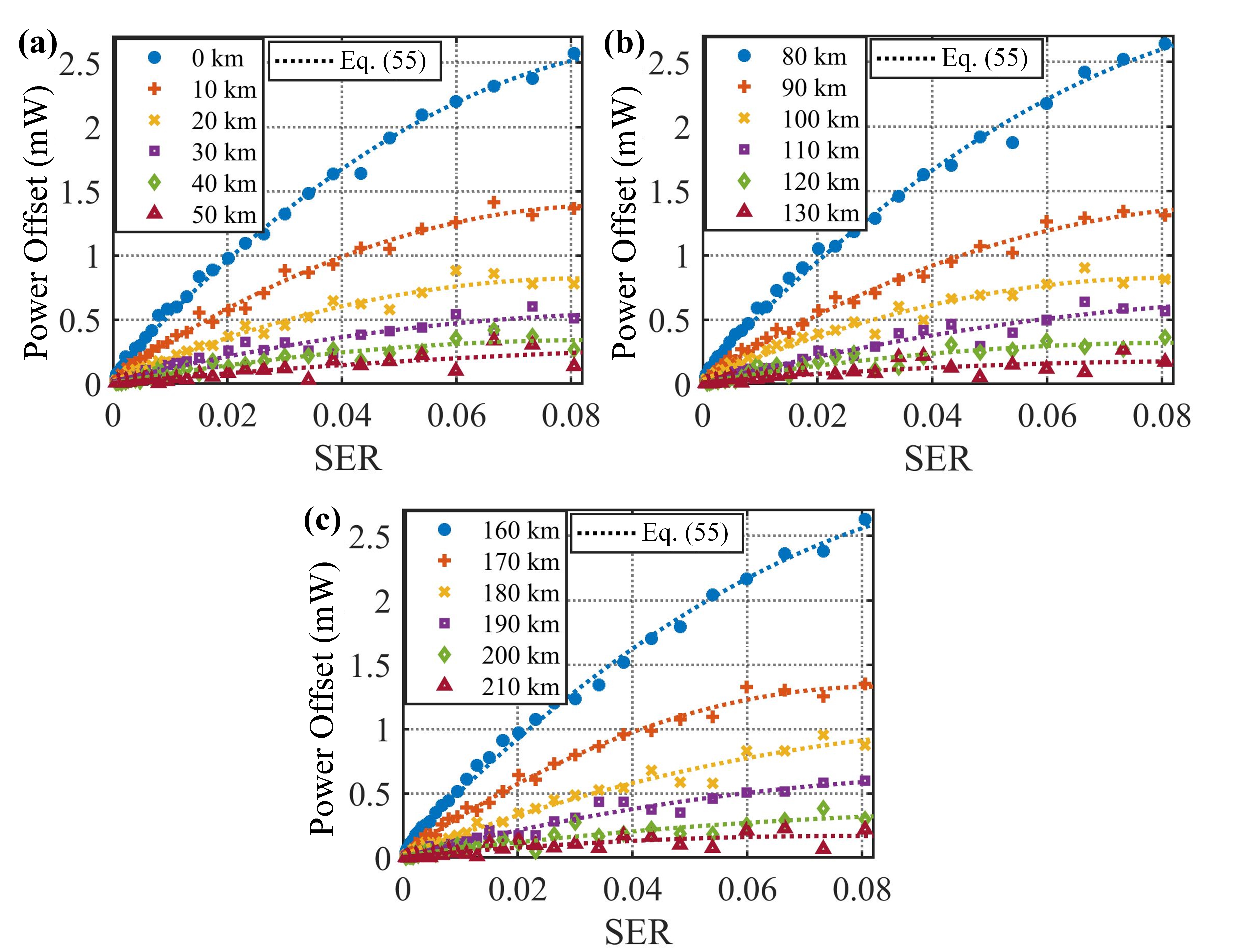}
\caption{The power offset in a unit of mW as a function of SER at the beginning of each fiber span (16-QAM with 8 dBm launch power). (a) 0 km - 40 km. (b) 80 km - 120 km. (c) 160 km - 200 km.}
\label{fig_8}
\end{figure}

\begin{figure}[t!]
\centering
\includegraphics[width=22 pc]{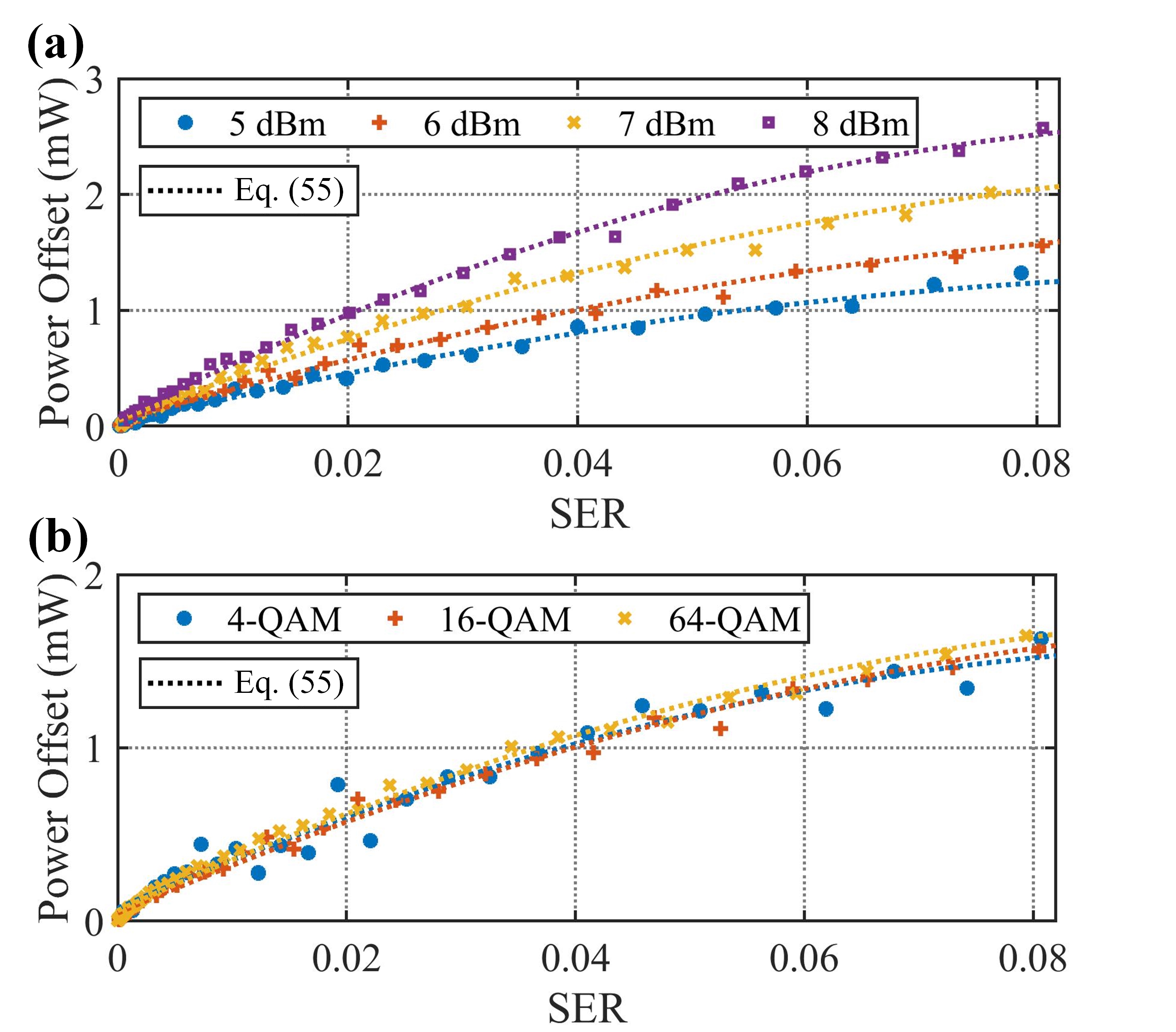}
\caption{The power offset in a unit of mW as a function of SER when $z=0$ km. (a) With different launch powers when the modulation format is 16-QAM. (b) With different modulation formats when the launch power is fixed at 6 dBm.}
\label{fig_9}
\end{figure}

Subsequently, we proceed to validate Eq. (55). We calculated the power offset at different positions according to $\boldsymbol{\widehat{\gamma^{\prime}}_{hd}}-\boldsymbol{\widehat{\gamma^{\prime}}_{tx}}$. It is important to note that a linear scale (measured in mW) is used here rather than a logarithmic scale (measured in dB), primarily because the power offset in Eq. (55) is expressed on a linear scale. Verification of the power offset in the logarithmic scale will be conducted subsequently. As illustrated in Fig. 6, when the modulation format is 16-QAM and the launch power is 8 dBm, the power offset varies at different positions and is notably higher at the beginning of each span than at the end, presenting a "sail shape". Moreover, the power offset increases with the SER, and this trend is especially pronounced at the beginning of each fiber span. The end of each fiber span is more vulnerable to stochastic noise due to the lower nonlinearity level mentioned above. Again, we focus on the beginning of each fiber span (within 40 km per span), and the power offset as a function of SER is shown in Fig. 7 (a)-(c). It should be noted that the legends in Fig. 7(a)-(c) indicate the measurement positions relative to the transmitter. For a given position, the scattered point represents the calculated power offset, and the dashed line represents the power offset according to Eq. (55). The scattered points fit well on the dotted line, clearly demonstrating that the nonlinear relationship between the power offset and SER aligns well with Eq. (55). Moreover, for a given SER, the power offset decreases as the measured distance within each span increases, which also indicates that the power offset depends on the distance due to the term $\gamma^{\prime}(n\Delta z)$ in Eq. (54), thereby initially verifying the correctness of Eq. (55).

Furthermore, to more comprehensively verify the correctness of Eq. (55), we also investigate the power offset as a function of SER under different launch powers and modulation formats. The measurement position of the power offset is set at 0 km to minimize the impact of noise. As shown in Fig. 8 (a), for a specific modulation format, e.g., 16-QAM, the nonlinear relationship between power offset and SER also adheres to $k\mathrm{SER}_{M-\mathrm{QAM},E_s} + p\sqrt{1-\mathrm{SER}_{M-\mathrm{QAM},E_s}}+q$ when the launch power varies from 5 dBm to 8 dBm. The reason for the distinct curves with different launch powers is also attributed to the term $\gamma^{\prime}(n\Delta z)$ in Eq. (54). Higher launch power corresponds to a higher $\gamma^{\prime}(n\Delta z)$, resulting in a greater power offset. Additionally, as depicted in Fig. 8 (b), the relationship in Eq. (55) still holds with different $M$-QAM modulation formats when the launch power is fixed at 6 dBm. The reason for choosing 6 dBm here is to ensure that the nonlinearity is not excessively severe, which would otherwise prevent the measured SER from reaching zero even when no noise (or a low noise level) is applied. Note that for different $M$ with the same launch power and overall SER, the theoretical $\left.\mathrm{SER}_{\sqrt{M}-\mathrm{ASK},E_s/2}\right|_{edge}$ in Eq. (43) varies, thus altering the constellation-related parameters in Eq. (49), resulting in slightly different curves for different modulation formats. 

\begin{figure}[t!]
\vspace{-0.4 pc}
\centering
\includegraphics[width=22 pc]{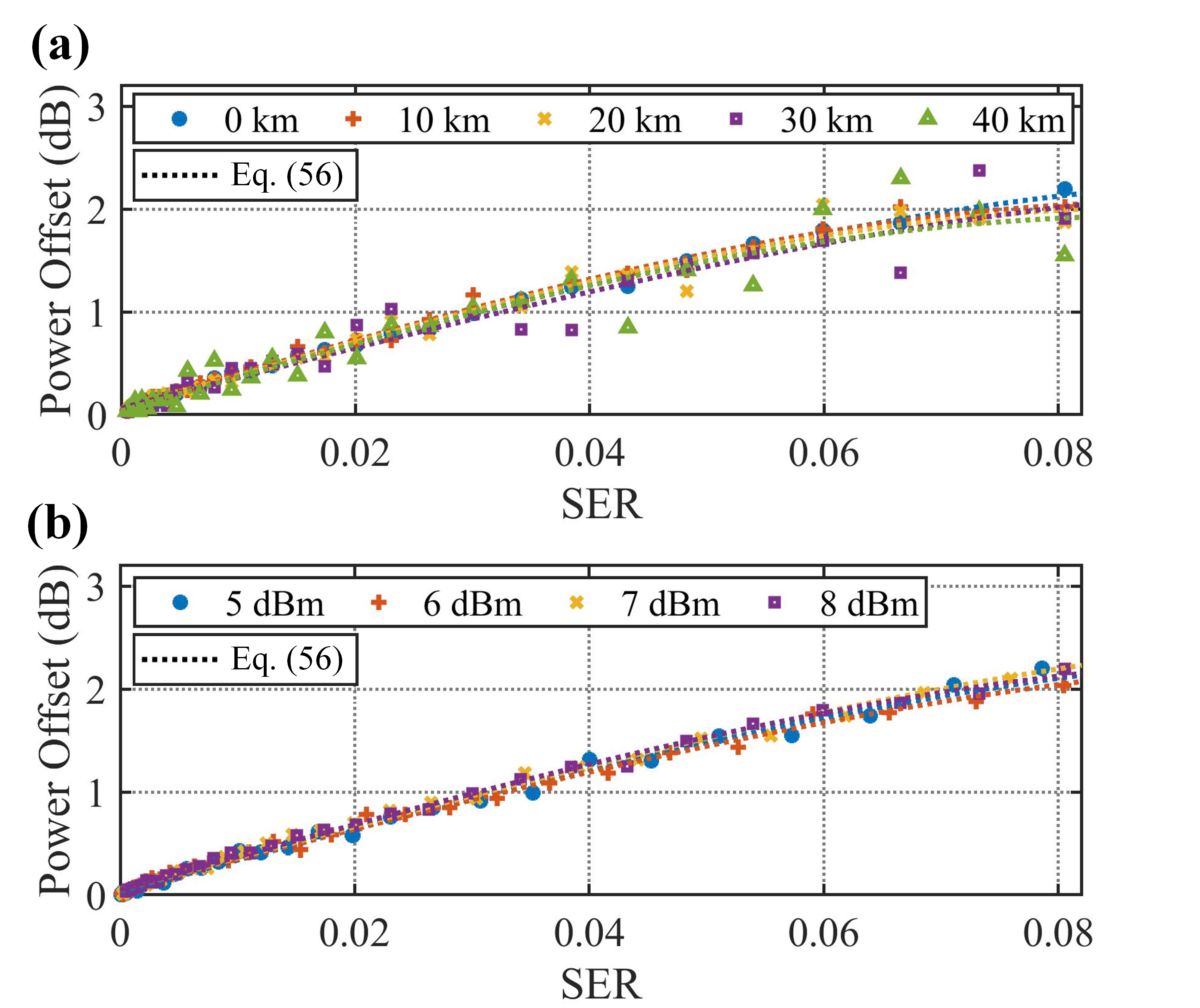}
\caption{The power offset in a unit of dB as a function of SER (16-QAM). (a) At different measurement positions when the launch power is fixed at 8 dBm. (b) With different launch powers when the measurement position is fixed at 0 km.}
\label{fig_1}
\end{figure}

Here, one may still be curious about the power offset in the logarithmic scale, as shown in \cite{T8}. The power offset in the logarithmic scale can be expressed as
\begin{equation}
\begin{aligned}
\mathrm{PO}_{\mathrm{dB}} &= 10\mathrm{log}_{10}(\boldsymbol{\gamma^{\prime}}) - 10\mathrm{log}_{10}(\boldsymbol{\widehat{\gamma^{\prime}}_{hd}})
=10\mathrm{log}_{10}\left(\frac{1}{1-\frac{\mathrm{PO}}{\boldsymbol{\gamma^{\prime}}}}\right)
\end{aligned}
\end{equation}
Then, the results of Fig. 7 (a) and Fig. 8 (a) after transforming the power offset from the linear scale to the logarithmic scale, are shown in Fig. 9 (a) and (b), respectively. In this instance, the power offsets at different measurement positions and launch powers are normalized to nearly the same level, due to the normalized term $\frac{\mathrm{PO}}{\boldsymbol{\gamma^{\prime}}}$ in Eq. (56). Under this logarithmic scale, the analytical dotted curves in Fig. 8 (a) converted according to Eq. (56) almost overlap. Since Eq. (56) is converted from Eq. (55), the relationship between the power offset in logarithmic scale and SER is now expressed as $10\mathrm{log}_{10}(1/(k\mathrm{SER}_{M-\mathrm{QAM},E_s} + p\sqrt{1-\mathrm{SER}_{M-\mathrm{QAM},E_s}}+q))$. This phenomenon is similar to the one revealed in \cite{T8} using commercial transceivers, although the authors in \cite{T8} did not provide a clear analytical explanation for the generation of power offset using HD data.

Through the numerical verification presented in this section, we have confirmed the following findings:

1) Power offset will be generated in MMSE-PPE if HD data is used. The analytical expression for the power offset can be obtained by introducing a virtual hard-decision nonlinear perturbation term.

2) The magnitude of power offset in MMSE-PPE using HD data is nonlinearly related to the SER of $M$-QAM modulation format, which can be expressed as $k\mathrm{SER}_{M-\mathrm{QAM},E_s} + p\sqrt{1-\mathrm{SER}_{M-\mathrm{QAM},E_s}}+q$.

\section{Possible future work of power offset mitigation}
This work primarily investigates the nature of power offset and its relationship to SER of $M$-QAM formats when MMSE-PPE is performed based on HD data. For other modulation formats, the relationship should be further verified. Moreover, the phenomenon of power offset needs to be taken seriously, especially in scenarios requiring absolute power estimation. The pre-FEC SER (or BER) is a standard telemetry parameter, usually estimated and given by commercial transceivers. According to the analytical relationship, one might design a power offset mitigation scheme directly: first, establish a look-up table of the power offset and SER, and then apply it in the actual implementation. The effectiveness of this kind of power offset mitigation process also requires further investigation.

\section{Conclusion}
This paper conducts sufficient theoretical derivation and analysis to explore the principle of power offset in MMSE-PPE caused by hard decision errors when pre-FEC HD data is used. By introducing a virtual hard-decision nonlinear perturbation term, we have provided the first analytical expression for the power offset in MMSE-PPE when using pre-FEC HD data. Enabled by this analytical expression, we have also established the first nonlinear relationship between the power offset and the SER of $M$-QAM formats under AWGN assumption. Through a numerical 130-GBaud single-wavelength coherent optical fiber transmission system with 4-QAM, 16-QAM, and 64-QAM formats under different SER situations, the correctness of the analytical expression of power offset and nonlinear relationship between the power offset and SER of $M$-QAM formats have been thoroughly verified. Overall, this theoretical investigation enriches the academic discourse on the power offset of MMSE-PPE using pre-FEC HD data, aiming to draw attention to this phenomenon, especially in scenarios requiring absolute power estimation. Moreover, this study also benefits the possible design of power offset mitigation in MMSE-PPE, advancing its potential for actual application.

\begin{backmatter}
\bmsection{Funding}
National Natural Science Foundation of China (62271080); The fund of State Key Laboratory of IPOC (BUPT) (IPOC2022ZT06).

\bmsection{Disclosures}
The authors declare no conflicts of interest.

\bmsection{Data availability} Data underlying the results presented in this paper are not publicly available at this time but may be obtained from the authors upon reasonable request.

\end{backmatter}


\bibliography{Optica-template}






\end{document}